# Table-top high-energy 7μm OPCPA and 260mJ Ho:YLF pump laser

U. Elu[1], T. Steinle[1], D. Sánchez[1], L. Maidment[1], K. Zawilski[2], P. Schunemann[2], U. D. Zeitner[3], C. Simon-Boisson[4], J. Biegert[1,5]

[1]ICFO - Institut de Cienciess Fotoniques, The Barcelona Institute of Science and Technology, 08860 Castelldefels, Barcelona, Spain.
[2]BAE Systems, MER15-1813, P.O. Box 868, Nashua, New Hampshire 03061, USA.
[3]Fraunhofer Institute of Applied Optics and Precision Engineering, A.-Einstein-Str. 7, 07745 Jena, Germany
[4]THALES Optronique S.A.S., Laser Solutions Unit, 2 avenue Gay-Lussac, 78995 Elancourt Cedex, France.
[5]ICREA, Pg. Lluis Companys 23, 08010 Barcelona, Spain.
*Corresponding author: ugaitz.elu@icfo.eu

**We present the state of the art of a compact high-energy mid-infrared laser system for TW-level 8-cycle pulses at 7 μm. This system consists of an Er:Tm:Ho:fiber MOPA which serves as the seeder for a ZGP-based OPCPA chain in addition to a Ho:YLF amplifier which is Tm:fiber pumped. Featuring all-optical synchronization, the system delivers 260-mJ pump energy at 2052 nm, 16-ps duration at 100 Hz with a stability of 0.8 % rms over 20 min. We show that chirp inversion in the OPCPA chain leads to excellent energy extraction and aids in compression of the 7-μm pulses to 8 optical cycles (188 fs) in bulk $BaF_2$ with 93.5 % efficiency. Using 21.7 mJ of the available pump energy, we generate 0.75-mJ-energy pulses at 7 μm due to increased efficiency with a chirp-inversion scheme. The pulse quality of the system's output is shown by generating high harmonics in ZnSe which span up to harmonic order 13 with excellent contrast. The combination of the passive carrier-envelope phase stable mid-infrared seed pulses and the high-energy 2052 nm picosecond pulses makes this compact system a key enabling tool for the next generation of studies on extreme photonics, strong field physics and table-top coherent X-ray science.**

Ultrafast and intense (>$10^{14}$ W/cm$^2$) pulses with carrier-to-envelope-phase (CEP) stability give rise to extreme photonic applications such as strong-field ionization [1, 2], or the generation of isolated attosecond pulses in the extreme ultraviolet range (XUV) [3, 4]. High-energy mid-infrared pulses through ponderomotive energy scaling provide unique capabilities for the generation of soft x-ray harmonics up to the keV level [5-10]. Intense mid-infrared pulses can be generated from compact table-top systems, providing an exciting approach towards a new generation of miniature particle accelerators though wakefield acceleration [11]. Low atmospheric losses at wavelengths between 8 – 14 μm provide an opportunity to exploit self-guided pulses via filamentation over kilometer distances without losing intensity through absorption or diffraction [1]. Thus, there is an important demand on high-energy mid-IR pulses which provide CEP controlled electric field waveforms.

The challenge of generating broadband mid-infrared pulses has been approached by direct laser generation, as well as difference frequency generation (DFG) or optical parametric amplification. Solid-state lasers based on Cr:ZnSe and Cr:ZnS have been demonstrated producing nJ-level short pulses in the 2 – 3 μm spectral range [12, 13], but amplification to the mJ-level has not been realized and no realistic materials exist to directly operate in mode-locked mode at mid-infrared wavelengths. DFG has been studied exploiting different nonlinear crystals, such as e.g. CSP, GaSe and OPGaP [14 – 16]. All these crystals show low-absorption coefficients in the mid-infrared as well as in the near-infrared (NIR) regime making them suitable to seed the DFG process with commercial NIR fiber lasers for the generation of mid-infrared pulses. Unfortunately, the generated mid-infrared pulse energy is typically limited to few pico-joules mainly due to the low damage threshold of these mid-infrared crystals and the high multi-photon absorption coefficients [17, 18]. Recent intrapulse DFG experiments [19] have generated broad mid-IR spectra with nJ-level pulse energy [20, 21] which are very promising for spectroscopy applications [22, 23] but do not reach ultra-intense levels.

Efficient energy scaling systems with suitable mid-infrared nonlinear crystals are therefore required to increase the energy to the mJ-level. Optical parametric chirped pulse amplification (OPCPA) can address this problem [24, 25], though it has been recognized that in order to amplify wavelengths beyond 5 μm a new pump platform had to be developed to optimize the quantum efficiency [26, 27] and mitigate crystal defect related damages [28]. Previously, we have demonstrated the first compact 7-μm mJ-level platform based on a Ho:YLF based chirped pulse amplifier (CPA) at 2 μm as the pump [26] and a ZGP-based OPCPA for the energy-scaling of the mid-infrared pulses [24] with all-optical synchronization [29]. However, using a diffraction grating-based pulse compressor in this spectral range at the end of an OPCPA restricts the pulse energy of the compressed output pulses due to the low efficiency of commercially available gratings.

Here we present a new scheme which incorporates a chirp inversion inside the OPCPA chain, thus permitting us to mitigate adverse chirp-related amplification effects and to use bulk material for stretching and compression. While the relative energy loss due to the insertion of the chirp inverter remains equal, placing it earlier in the amplification chain reduces the absolute energy spent for dispersion tailoring and thereby increases the overall system conversion efficiency. Thus, the overall efficiency of the OPCPA after compression could be doubled with the chirp inversion scheme. We have also upgraded the pump laser system from 40 mJ to 260 mJ at 100 Hz repetition rate, whilst maintaining the high stability and reliability of the CPA system. The high-energy 2052-nm pump laser is used for amplification of the optically-synchronized passively CEP stable 7-μm seed pulses in an OPCPA chain separated by the chirp inverter, analog to the previous demonstration with our 3 μm OPCPA system [30]. We start by discussing the 2052-nm CPA pump laser system and then report on our results in energy-scaling and characterization of mJ-level 7 μm femtosecond pulses.

The conceptual layout of the high-energy Ho:YLF CPA is shown in Fig. 1. Narrowband nJ pulses from the Er:fiber laser are temporally stretched to 340 ps by a chirped volume Bragg grating (CVBG) and the repetition rate is reduced to 100 Hz with a Pockels cell pulse picker. The 100-Hz pulses are then used to seed a regenerative amplifier based on a water-cooled Ho:YLF crystal, which is pumped by 24 W from a CW Tm:fiber laser (IPG Photonics) centered at 1940 nm. The regenerative amplifier achieves amplification from 2 nJ to 4-mJ pulse energy. Its cavity is operated in saturation ensuring highly stable performance and excellent beam quality, both being crucial for optimum performance of the entire system. Subsequently, the 4-mJ pulses are introduced in a booster multi-pass amplifier for energy scaling purposes. The booster amplifier is based on a loose focusing triple-pass geometry using a 50-mm-long Ho:YLF active material which is cryogenically-cooled to 95 K. The booster is pumped in a single-pass loose focusing geometry using a 120-W CW Thulium fiber laser (IPG Photonics) centered at 1940 nm. The booster amplifier increases pulse energy from 4 mJ to 42 mJ in the first pass, to 173 mJ in the second pass, and to 260 mJ in the third pass.

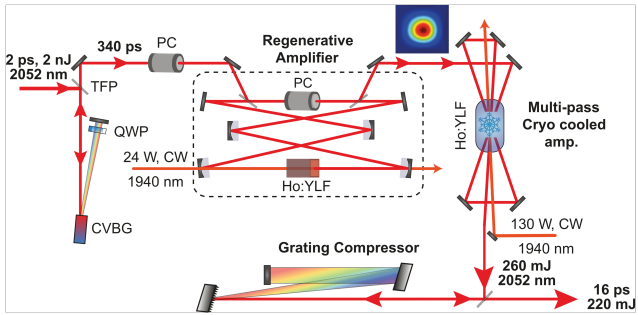

Fig. 1. (a) Layout of the 2052 nm Ho:YLF solid state laser system. The narrowband 2052-nm pulses from the three-color fiber front-end are stretched in a chirp volume Bragg grating (CVBG) and the repetition rate is reduced in a Pockels cell (PC) before amplification in a water-cooled regenerative amplifier and cryogenic cooled booster amplifier consecutively. Finally, the stretched pulses are back compressed using a pair of dielectric coated gratings.

The pulse-to-pulse stability of 0.8% rms over 20 min at 260 mJ (26 W) is shown in Fig. 2(a). Together with excellent beam quality this makes the system ideal for pumping the mid-infrared OPCPA chain.

To temporally compress the high energy 2052-nm pulses the beam size is increased to 6 mm (FWHM) to avoid damage to optics. The grating compressor consists of two dielectrically coated gratings (Fraunhofer IOF), with 900 lines/mm and has a footprint of 140 cm by 25 cm. We measure an overall efficiency of 85 % corresponding to a final compressed pulse energy of 220 mJ.

Figure 2(b) shows results from the characterization of the high-energy picosecond pulses via a non-collinear intensity autocorrelation using a 200-μm-thick BBO crystal for second harmonic generation. The pulse duration, assuming a Gaussian deconvolution is 16 ps; the wings results from uncompensated higher order phase. While the 2052-nm spectrum allows pulse compression down to 5 ps, we set the pulse duration to 16 ps in order to maintain a reasonable peak intensity within the limited aperture on the ZGP nonlinear crystals for the OPCPA stages.

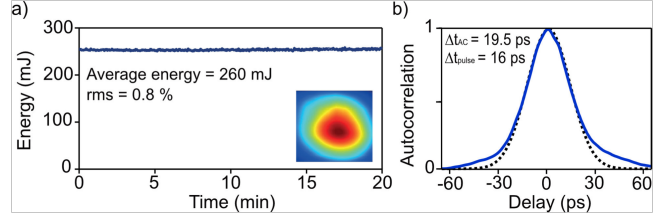

Fig. 2. (a) Measured stability at full power from the Ho:YLF pump system. The inset shows the beam profile measured at full power. (b) Measured intensity autocorrelation (blue) of the 2052-nm pump pulses after compression in the dielectrically-coated gratings together with a Gaussian fit (dashed black). The FWHM duration of the autocorrelation signal was measured to be 19.5 ps corresponding to a pulse duration of 16 ps at FWHM.

Figure 3 shows a schematic of the 7 μm OPCPA system. The two femtosecond outputs from the Er:Tm:Ho:fiber frontend are combined in collinear geometry in a 1-mm-thick GaSe nonlinear crystal for type I DFG. Mixing the 1.5 μm, 70 fs and 2 μm, 100-fs pulses in GaSe, we are able to generate 7 pJ of mid-infrared pulses centered at 7 μm.

The 7μm DFG pulses are stretched in a 16-mm-long $BaF_2$ rod to a pulse duration of 4 ps. All the OPCPA stages are based on non-collinear geometry with external angle between pump and seed of 6.5 degrees and type I phase-matching configuration in ZGP nonlinear crystals (BAE Systems). The OPCPA consists of two sections, a pre-amplification and a booster-amplification section with an intermediate chirp inversion stage. The pre-amplification section is based on two 3-mm-thick uncoated ZGP-based OPCPA stages pumped by 200 μJ and 1.5 mJ 2052-nm pulses and achieving 7 μm amplification to 0.1 μJ and 20 μJ respectively.

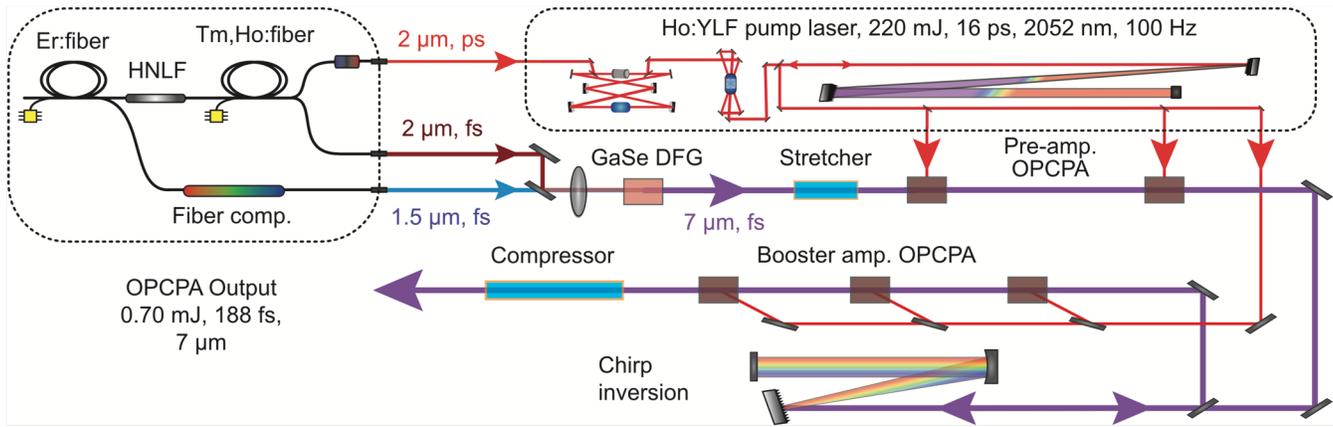

Fig. 3. Layout of the 7 µm OPCPA. The mid-infrared seed is generated using the two broadband femtosecond outputs from a three-color fiber front-end via DFG. Afterward, the mid-infrared pulses are stretched in a dielectric bulk and consecutively amplified in a pre-amplifier and a booster-amplifier separated with a chirp-inversion stage. Maximum efficiency of the OPCPA is achieved by tailoring the seed-to-pump pulse durations in the pre and booster amplifiers. The broadband high-energy mid-infrared pulses are recompressed using a dielectric bulk rod of $BaF_2$.

The chirp inversion stage between the amplification sections is based on a Martinez-type stretcher which provides control on the stretching factor of the 7-µm pulses, optimizing the efficiency of the booster OPCPA stages and allowing the final pulse compression in a bulk dielectric medium. The chirp inverter is aligned to stretch the 7-µm pulses from a negatively chirped 4-ps duration ($-1.77 \times 10^5$ fs$^2$) to positively chirped 11-ps pulse duration ($5.3 \times 10^5$ fs$^2$) in order to optimize the temporal overlap between mid-infrared seed and 2052-nm pump pulses. Moreover, the chirp inverter does not only give us the opportunity to tailor the seed pulse duration, but also ensures an efficient compression scheme in dielectric bulk at the end of the OPCPA. This is important as in the current implementation the chirp inversion stage exhibits an overall efficiency of 10 % due to the limitations imposed by commercially available mid-IR gratings. The chirp inverter was placed in the pre-amplification stages of the OPCPA after reaching an acceptable power level at 7 µm for the alignment of the corresponding optics.

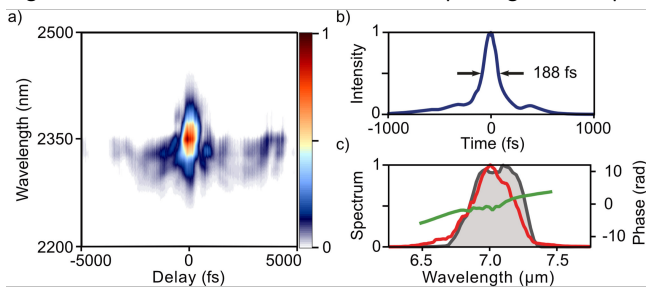

Fig. 4. Pulse characterization of the 7 µm compressed pulses, a) the measured third harmonic FROG, b) the retrieved pulse envelope with 188 fs FWHM duration and c) measured (filled grey) and retrieved spectrum (red line) and phase (green line).

The booster amplifier is based on three consecutive OPCPA stages, with the first stage based on 5-mm-thick uncoated ZGP largely to compensate the losses imparted by the chirp inverter, to restore the pulse energy to 50 µJ. The next and the last stage are comprised by two identical 2-mm-thick anti-reflection (AR) coated ZGP crystals and pumped by 3.5 mJ and 15 mJ. This configuration results in amplified pulse energies of 0.25 mJ and 0.75 mJ, respectively. The pump peak intensity in the OPCPA stages was set to be between 2-4 GW/cm$^2$ to avoid damage on the ZGP crystals. The pump energy used in the OPCPA is presently limited to 21.7 mJ due to the aperture size of 8 mm x 8 mm of the ZGP crystals. Due to the challenges associated with proper characterization and propagation of the mid-infrared pulses, all the linear and nonlinear processes were numerically pre-simulated using the Simulation System for Optical Science (Sisyfos) nonlinear wave propagation code [31].

The final compression of the mid-infrared pulses occurs in a 48-cm-long $BaF_2$ rod that compensates the dispersion acquired in the chirp inversion stage and shows an overall transmission efficiency of 93.5 % corresponding to a final compressed pulse energy of 0.7 mJ. Characterization of these pulses is done using a home-built all-reflective-optics-based third harmonic generation frequency-resolved optical gating (THG FROG) device using a thin ZnSe crystal for the generation of the third harmonic. This characterization technique allows us to measure the signal of the FROG using a thermo-electric cooler (TEC) based NIR spectrometer without the need of using mid-infrared nitrogen cooled detectors or less precise mid-infrared photodiodes. Furthermore, the THG FROG has allowed a fast reconstruction of full pulse amplitude and phase of our high-energy mid-infrared pulses showing a FWHM pulse duration of the 7-µm pulses of 188 fs, see Fig. 4.

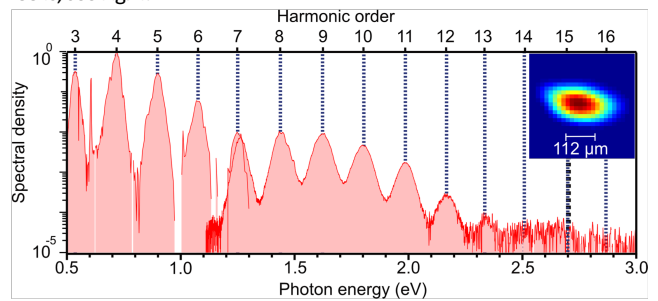

Fig. 5. Measured high harmonic spectrum in bulk ZnSe crystal (red curve) and the harmonic orders from 3rd to 13th harmonic, both even and odd, shown in dashed blue lines. The black thick dashed line refers to the band gap of the crystal. The inset shows the measured beam profile of the 7 µm pulses at the generation plane of the harmonics.

To test the fidelity of the compression we employed the high-energy broadband mid-infrared pulses for high harmonic generation (HHG) in a ZnSe sample. Owing to the materials' polycrystalline nature, we observe both even and odd harmonics up to the 13th order (Fig. 5). The high-energy 7-µm pulses are focused in the sample to a spot size of 136 µm at FWHM, which is shown in the inset of Fig. 5, using a 75 mm off-axis parabolic mirror and the visible and NIR harmonics are then collimated with a f = 75 mm fused silica lens. Using these focusing conditions and employing 50 µJ of 7-µm energy (limited to avoid damage to the ZnSe) we estimate the peak intensity to be 0.45 TW/cm$^2$. By splitting the NIR and visible spectral parts using a silicon thin plate, we measure all the harmonics below the band gap simultaneously with integration times lower than 500 ms. The HHG emission is fiber-coupled into the spectrometers using a NIR256 spectrometer (Ocean Optics) for the spectral range of 900 nm to 2550 nm and a HR4000 visible spectrometer (Ocean Optics) for the range of 200 nm to 1100 nm.

In conclusion, we have demonstrated an all-optically synchronized CPA and OPCPA system generating 260-mJ pulses centered at 2052 nm and 0.75 mJ mid-infrared pulses at 7 µm at 100 Hz repetition rate. We demonstrated that the chirp inversion scheme improved the total OPCPA efficiency and allowed reaching the mJ-level with a demonstrated pulse energy of 0.7 mJ at 188 fs after compression (8 optical cycles). The quality of the compressed high-energy 7-µm pulses was demonstrated through HHG in ZnSe by achieving a spectrum expanding over all the visible and NIR regime up to order 13. This mid-infrared OPCPA platform features the highest demonstrated energies for both 2052-nm and 7-µm pulses and represents a significant step towards mid-infrared high-field applications such as coherent keV x-rays or self-similar atmospheric propagation and laser-plasma acceleration.

**Acknowledgments** Funding was gratefully provided by the Spanish Ministry of Economy and Competitiveness for Plan Nacional FIS2017-89536-P, the "Severo Ochoa" Programme for Centres of Excellence in R&D (SEV-2015-0522), the European Union's Horizon 2020 programme for Laserlab-Europe (654148), FET-OPEN "PETACom" (829153), the Catalan Agencia de Gestió d'Ajuts Universitaris i de Recerca (AGAUR) with 2017 SGR 1639 Fundació Cellex Barcelona the CERCA Programme / Generalitat de Catalunya., Army Research Laboratory (W911NF-17-1-0565), ERC Advanced Grant "TRANSFORMER" (788218) and ERC Proof-of-Concept Grant "miniX" (840010).